\documentclass[aps,prl,reprint,groupedaddress,nofootinbib]{revtex4-1}
\usepackage{amsmath}
\usepackage{amssymb}
\usepackage{bbm}
\usepackage{bm}
\usepackage{color}
\usepackage{booktabs} 
\usepackage{tikz}
\usepackage{cancel}

\newcommand*\diff{\mathop{}\!\mathrm{d}}

\usepackage{hyperref}

\hypersetup{
colorlinks=true,         
linkcolor=blue,          
citecolor=red,        
urlcolor=red            
}

\renewcommand{\vec}[1]{\mathbf{#1}}

\begin{document}

\title{Light Cone in a Quantum Spacetime}

%

\author{Flavio~Mercati}
\email[]{flavio.mercati@gmail.com}
\affiliation{Dipartimento di Fisica, Sapienza Universit{\`a} di Roma,\\
P.le A. Moro 2, 00185 Roma, Italy.}

\author{Matteo~Sergola}
\email[]{matteo.sergola@gmail.com}
\affiliation{Dipartimento di Fisica, Sapienza Universit{\`a} di Roma,\\
P.le A. Moro 2, 00185 Roma, Italy.}

\date{\today}

\begin{abstract}
Noncommutative spacetimes are a proposed effective description of the low-energy regime of Quantum Gravity. Defining the microcausality relations of a scalar quantum field theory on the $\kappa$-Minkowski noncommutative spacetime allows us to define for the first time a notion of light-cone in a quantum spacetime. This allows us to reach two conclusions. First, the majority of the literature on $\kappa$-Minkowski suggests that this spacetime allows superluminal propagation of particles. The structure of the light-cone we introduced allows to rule this out, thereby excluding the possibility of constraining the relevant models with observations of in-vacuo dispersion of Gamma Ray Burst photons.
Second, we are able to reject a claim made in [\href{https://journals.aps.org/prl/abstract/10.1103/PhysRevLett.105.211601}{Phys. Rev. Lett. 105, 211601 (2010)}], according to which the light-cone of the $\kappa$-Minkowski spacetime has a `blurry' region of Planck-length thickness, independently of the distance of the two events considered. Such an effect would be hopeless to measure. Our analysis reveals that the thickness of the region where the notion of timelike and spacelike separations blurs grows like the square root of the distance. This magnifies the effect, \emph{e.g.} in the case of cosmological distances, by 30 orders of magnitude.
\end{abstract}

\pacs{}

\keywords{quantum gravity phenomenology, light cone,  microcausality, noncommutative spacetimes,  Gamma Ray Bursts, high-energy neutrinos}

\maketitle

\section{Introduction}

One of the most studied phenomenological windows on Quantum Gravity (QG) is the possibility that spacetime acquires a `fuzzy' or `grainy' quality at small scales~\cite{Wheeler1955,Hawking1978,Nick_foam1,Nick_lightcone}. This is believed to have an effect on the propagation of particles through vacuum -  spacetime itself is supposed to behave like a dispersive medium. Such conjectured effects, although expected to be minuscule,\footnote{The expectation is that the scale controlling these effects is the Planck length, $L_p \sim 10^{-35} m$, or  the associated energy $E_p \sim 10^{28} eV$.} could accumulate over cosmic distances to such an extent that they may become observable. In fact, it turns out that the magnitudes of the quantities at our disposal combine in a fortunate way for a particular kind of astrophysical sources: Gamma Ray Bursts (GRBs). These sources can be as far as several billion light years, and emit particles which can reach energies of hundreds of GeVs. 
Suppose that an effect controlled by the Planck length follows a law of the kind (in $\hbar = c =1$ units)
\begin{equation}\label{planckeffect}
\Delta t = L_p \, E \, L \,,
\end{equation}
where $E$ is the energy of the particle, $L$ is the distance of the GRB and $\Delta t$ is the time scale of the conjectured effect. Replacing the numbers   $E \sim 1 ~ TeV$ and $L = 10^9 ~ ly$, we get $\Delta t \sim 2 ~ s$, which is quite macroscopic.
This coincidence of relevant scales is an invitation to use GRB to constrain new physics whose characteristic scale is $E_p$, and which might have a quantum-gravitational origin. 

The most studied proposals conjecture that, due to the fuzziness of the background spacetime, the relativistic kinematics of particles of very high energy departs from what predicted by Special (or General) Relativity. 
Their dispersion relation is modified by corrections controlled by the Planck scale~\cite{21,MagueijoSmolin2002,MagueijoSmolin2003,GiovanniLRR,Nick_string_refractive,MagueijoSmolin_string_dsr}. These effects may be \emph{systematic,} \emph{i.e.} for a massless particle:
\begin{equation}
 v = 1 + k \, L_p \, E  +  \mathcal O (L_p^2) \,, \label{SystematicDispRel}
\end{equation}
so each particle of energy $E$ goes faster (or slower) than light by a quantity $|k| \, L_p \, E$,
where $k \in \mathbbm{R}$. Another possibility is that the effects are  \emph{statistical,} in the sense that the average speed of massless particles is still $1$, but there are fluctuations which give a variance:
\begin{equation}
\Delta v =  |k| \, L_p \, E  +  \mathcal O (L_p^2) \,. \label{StatisticDispRel}
\end{equation}
This implies that particles originated at a point-source and propagating through vacuum will arrive with a time spread [the $\Delta t$  of Eq.~\eqref{planckeffect}]  of order $|k| \, L_p \, E$.

In any case, one can hope to put experimental constraints on the effects formalized by relations like~(\ref{SystematicDispRel}) and (\ref{StatisticDispRel}) with GRB observations  (see for example references \cite{21} and \cite{Vasileiou2015}, which provide bounds on formulas of such kind), precisely because of the coincidence in the relevant scales we observed above. GRBs are not pointlike sources (neither in space nor in time), but at the source they are known to be of reasonably short duration. In particular, about 30\% of the GRBs are classified as `short', with a duration of less than 2 seconds, and with recognizable features (`pulses') in their light curve that last a fraction of a second~\cite{Fermi2009Nature,Nick_GRB_finesse_analysis}. This allows to put a limit to $|k|$ in Eqs.(\ref{StatisticDispRel}) and~(\ref{SystematicDispRel}): for photon energies in the range  $0 - 100 ~ GeV$, simultaneous detection implies $|k|\lesssim 100$. GRB photons are not the only possible way to test these effects: they are now being tested~\cite{Amelino-Camelia2017}  with very high-energy ($\sim 100 ~ TeV$) neutrino detections of the IceCube telescope~\cite{Aartsen2013,Aartsen2014}, with gravitational waves~\cite{Nick_gravity_waves}, and with cosmological observations~\cite{Amelino-Camelia2013a}.
The analyses that put constraints on effects like~(\ref{SystematicDispRel}) and (\ref{StatisticDispRel}) have now reached a high degree of sophistication, constraining $k$ down to $\mathcal O (1)$~\cite{Nick_GRB_finesse_analysis,Fermi2009Nature,Xu2016,Xu2016a,Amelino-Camelia2017}.


On the theoretical side, however, most studies did not go much further than conjecturing more or less arbitrarily a relation of the kind~(\ref{SystematicDispRel}) or (\ref{StatisticDispRel}) -- not without making heavy compromises on the rigor.
For example, it is widely expected that certain models of quantum geometry  called \emph{noncommutative spacetimes} predict in-vacuo dispersion of the form~(\ref{SystematicDispRel}) or~(\ref{StatisticDispRel}). These theories model the `fuzzy' nature of spacetime at the Planck scale as some  commutation relations for the coordinates (which are promoted to quantum operators $\hat x^\mu$): $[\hat x^\mu, \hat x^\nu] = i \, \Theta^{\mu\nu}(\hat x)$, for some given function $\Theta^{\mu\nu}(\hat x)$~\cite{doplicher1995,NikitaNekrasov}. The uncertainty relations that follow
from such commutators $\Delta x^\mu \Delta x^\nu \geq  \frac 1 2 \langle \Theta^{\mu\nu}(\hat x) \rangle$ impose limits to the simultaneous measurability of multiple coordinates~\cite{doplicher1995}. Part of the reasons why noncommutative spacetimes are interesting is that, in 2+1 dimensions they arise as the low-energy limit of QG~\cite{Matschull,freidel}.

Perhaps the best-known example of noncommutative spacetime is \emph{$\kappa$-Minkowski}\footnote{Apart from being a popular model of noncommutative geometry, $\kappa$-Minkowski and its symmetry algebra are (almost) the only choices in 4 spacetime dimensions~\cite{noi2}.}~\cite{lukkkk,mr,Benedetti2009}. In this model, the  Poincar\'e algebra is deformed into a \emph{Hopf algebra}~\cite{compact,104,Landi:1997sh}, such that the commutation relations become nonlinear functions of the generators.\footnote{Something that is not allowed in Lie algebras.}
The $\kappa$-Poincar\'e Casimir is not a quadratic function of momentum anymore. In an expansion in powers of $L_p$:
\begin{equation} \label{DefCasimir_naive}
\mathcal C' = - E^2 + |\vec{p}|^2 + k \, L_p \, E^3 + \mathcal O(L_P^2) \,,
\end{equation}
Then, it is argued, the group velocity of a wave respecting $\mathcal C' = 0$,  will  be $v = \frac{\partial E}{\partial p} =  1  + k \, L_p \,  E$, precisely as in Eq.~(\ref{SystematicDispRel})~\cite{amelinowaves}.

This simple argument, however, is rather naive: the form~(\ref{DefCasimir_naive}) of the Casimir depends on the choice of basis of the $\kappa$-Poincar\'e algebra, and any nonlinear redefinition of basis is allowed in a Hopf algebra. By deducing a dispersion relation from the Casimir of $\kappa$-Poincar\'e, one ends up with a \textit{different} physical prediction for each choice of basis.

A more rigorous strategy  is to use Quantum Field Theory (QFT) on $\kappa$-Minkowski. Reference~\cite{Brasiliani} is highly significant in this context: by analyzing the retarded and advanced Green functions the authors are able to study their causality properties. Their conclusions are diametrically opposite to those of the advocates of~(\ref{SystematicDispRel}) or~(\ref{StatisticDispRel}): the light-cone is blurred over a region of Planckian size, independently of the  distance from the origin. This conclusion would make the microcausality properties of $\kappa$-Minkowski hopelessly untestable, because the departures from commutative Minkowski spacetime manifest themselves only at Planckian scales.
This conflicts with the intuition provided by the commutation relations of $\kappa$-Minkowski shown below in~(\ref{k-Minkowski_algebra}), which imply uncertainty relations of the form $\Delta z^1 \Delta z^0 \geq {\frac 1 2} L_p \langle z^1 \rangle$. If the spatial coordinate $z^1$ and the temporal one $z^0$ have uncertainties of the same order of magnitude, they will be of the order of $\sqrt{L_p \langle z^1 \rangle}$.\footnote{This was first observed in~\cite{Giovanni-kappa-LimitsMeasurability}.} For macroscopic distances, this is a colossal boost in the size of the effect, with respect to what was proposed in~\cite{Brasiliani}. 


The analysis of~\cite{Brasiliani} is definitely a step further from arbitrarily postulating a relation of the kind~(\ref{SystematicDispRel}) or (\ref{StatisticDispRel}), but it  relies on an inadequate assumption: the deformed Casimir of $\kappa$-Minkowski is simply promoted to a nonlinear wave operator on a space of \emph{commutative} functions, and its Green functions are calculated by inverting classical partial differential equations.
Again, this approach will yield different results in different Hopf-algebra bases.

In~\cite{Mercati2018} we  developed a framework that allows to discuss causality issues in a more realistic setting: a free Klein--Gordon field theory on $\kappa$-Minkowski. Our construction is Hopf-algebra-basis independent. The commutation relations between fields at different spacetime points were defined through an appropriate generalization of the Pauli--Jordan function, which is covariant under the action of the $\kappa$-Poincar\'e Hopf algebra. In commutative spacetimes the Pauli--Jordan function is zero outside of the light-cone, and  represents a useful field-theoretical definition of the causal structure. 
In the following analysis we retain the interpretation of the Pauli--Jordan function as a the commutator between fields at different spacetime points, establishing thereby the causal structure of spacetime. Even though we believe this is a reasonable assumption we stress that it should only be considered as such, and that our proposal should be further explored and supported by different calculations and examples.

In this Letter we extract the causal structure of $\kappa$-Minkowski from our noncommutative Pauli--Jordan function. We have multiple goals, the completion of which has a significant impact over a large literature in QG-phenomenology and noncommutative QFT:
\begin{enumerate}
\item Putting under scrutiny the claim of Ref.~\cite{Brasiliani} that the lightcone of $\kappa$-Minkowski is blurred over a region of invariably Planckian size. An improvement like what is suggested by the  $\kappa$-Minkowski uncertainty relations would be extremely significant, making the nonlocal effect potentially measurable.

\item Updating the discussion of superluminal propagation in  $\kappa$-Minkowski. If in-vacuo dispersion relations of the form~(\ref{SystematicDispRel}) or (\ref{StatisticDispRel}) is indeed predicted by QFT on $\kappa$-Minkowski, as appears to be commonly accepted in the literature~\cite{amelinowaves,Aloisio:2004yz}, then we already have bounds of Planckian magnitude on the relevant models~\cite{Nick_GRB,Fermi2009Nature,Xu2016,Xu2016a,Amelino-Camelia2017}. On the other hand, if the light-cone turned out to be blurred over a region of the same size as $\sqrt{L_p \langle z^1 \rangle}$ or less, then we couldn't talk about in-vacuo dispersion, because it would be forbidden by microcausality relations. GRBs would not put constraints on $\kappa$-Minkowski models.
\end{enumerate}

\section{QFT on $\kappa$-Minkowski}

The algebra of functions on $\kappa$-Minkowski in $n$ spatial dimensions $\mathcal{A}$ is generated by the $n+1$ coordinates $\hat z^\mu$:
\begin{equation}\label{k-Minkowski_algebra}
[\hat z^0 , \hat z^i] = \frac{i}{\kappa} \hat z^i \,,~~
[\hat z^i , \hat z^j] =0 \,, \qquad \hat z^\mu \in \mathcal A \,,
\end{equation}
The eponymous energy constant $\kappa$ is supposed to be related to the Planck scale.
In~\cite{Mercati2018} we introduced the Pauli--Jordan function for a complex scalar as an element of $\mathcal A \otimes \mathcal A$: the algebra of functions of two variables, generated by the operators $\hat x^\mu =\hat z^\mu \otimes 1$ and $\hat y^\mu = 1 \otimes\hat z^\mu$. The commutator between a quantum scalar field $\hat \phi(\hat x)$ and its conjugate $\hat \phi^\dagger (\hat y)$  at different spacetime points was defined as $[ \hat \phi(\hat x) , \hat \phi^\dagger (\hat y) ] =
i \hat{\Delta}_\text{PJ}(\hat x,\hat y) $, where:
\begin{equation}\label{Noncommutative_Pauli-Jordan_3D}
\begin{aligned}
&i \hat{\Delta}_\text{PJ}
= -
\int_{\mathbbm{R}^n} \!\!\!\!\!  \diff^n p \frac{e^{3 \frac{\omega^-  }\kappa}  e^{i {\vec p} \cdot \hat{\vec x}} e^{- i  e^{\frac{\omega^-  }\kappa} {\vec p} \cdot \hat{\vec y}}  e^{i \omega^-  (\hat{x}^0 - \hat{y}^0)}
}{ 2 \sqrt{ \left(m^2+| \vec p |^2\right)}} +
\\
&\int_{\mathbbm{R}^n} \!\!\!\!\!  \diff^n p 
 \frac{
\text{sign}(\kappa-|\vec p|) e^{3 \frac{\omega^+  }\kappa}  e^{i {\vec p} \cdot \hat{\vec x} - i e^{\frac{\omega^+  }\kappa} {\vec p} \cdot \hat{\vec y}  }  e^{i \omega^+  (\hat{x}^0 - \hat{y}^0) } }{ 2 \sqrt{ \left(m^2+| \vec p |^2\right)}}
\,.
\end{aligned}
\end{equation}
where
\begin{equation} \textstyle
\omega^\pm   = - \frac{\kappa}{2} \log \left[ 1 + \frac{2 m^2 + | \vec p |^2}{\kappa^2} \mp \frac{  \sqrt{\left(\kappa ^2+m^2\right) \left(m^2+| \vec p |^2\right)}}{\kappa^2} \right]\,,
\end{equation}
are the \emph{frequencies}, which, for $m \ll \kappa $, tend to the usual  $\pm\sqrt{\vec{p}^2+m^2}$ in the low-momentum limit $| \vec p | \ll \kappa $ .

 \begin{figure*}[t!]
\begin{center}
\raisebox{-1pt}{\includegraphics[height=0.195\textheight]{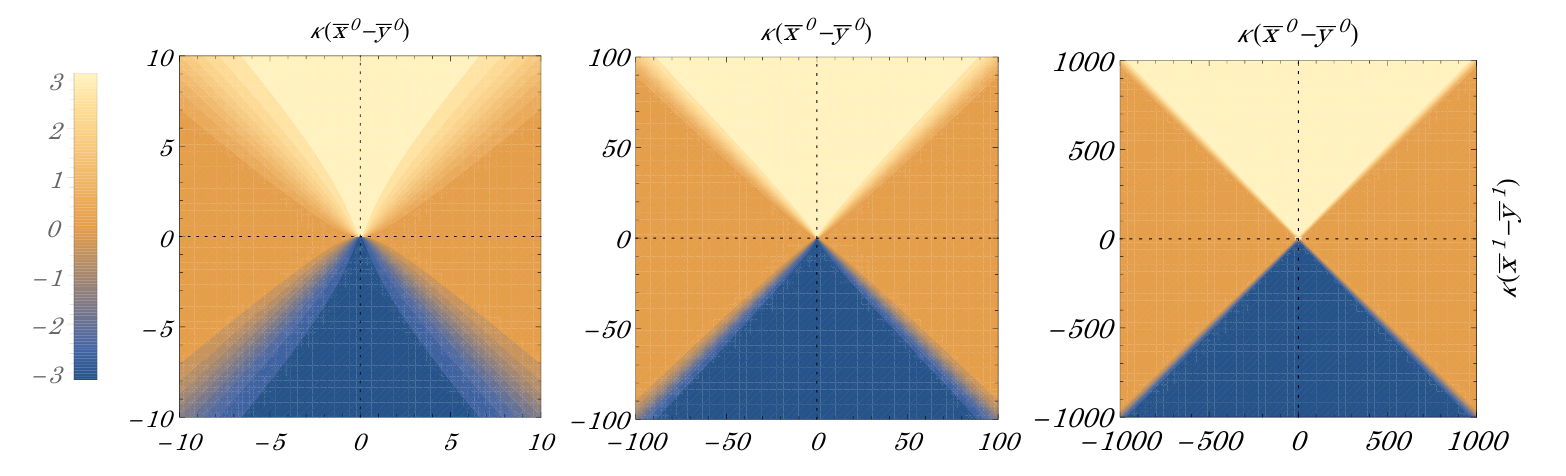}}
\end{center}\caption{Expectation value of the noncommutative Pauli--Jordan function calculated on a semiclassical state centred on the points $\bar x^\mu$,  $\bar y^\mu$. The function is represented as a contour plot on the plane $(\bar x^1 -\bar y^1)$-vs.-$(\bar x^0 -\bar y^0)$. All coordinates are expressed in Planck units. The diagram on the left is zoomed in at the origin, and shows a region of $20\times20$ Planck units $\kappa^{-1}$. The central diagram shows $200\times200$ Planck units, while the one on the left shows $2000\times2000$. It is apparent how, zooming out up to macroscopic distances form the origin, the noncommutative Pauli--Jordan function is indistinguishable from the commutative one, which is $\pi$ in the absolute future, $-\pi$ in the absolute past and $0$ in the absolute present. Compare these figures with the one shown in~\cite{Brasiliani}, where the `blurry' region has the same thickness everywhere.
An animated version of the diagram above is available at~\cite{Weblink}.
}\label{LightCone2D_1}
\end{figure*}

\section{Noncommutative light-cone}

Our present goal is to extract a commutative function of two variables from the much richer structure of the noncommutative function~\eqref{Noncommutative_Pauli-Jordan_3D}. To do so we take the expectation value of $\hat{\Delta}_\text{PJ}(\hat x,\hat y)$ on a \emph{state} on the 
algebra generated by $\hat x$, $\hat y$. In fact, having promoted spacetime coordinates to noncommuting operators, we can interpreted them as operators acting on a Hilbert space. The latter is a space of wavefunctions of the coordinates, which encode the probability of finding a certain value for the spacetime coordinates, and will respect uncertainty relations that prevent the points to be sharply localized. Now, $\kappa$-deformed quantum field theory is an effective theory, supposedly valid in the low-energy regime of a more fundamental quantum theory of gravity. Therefore it cannot be expected to predict these wavefunctions: the state of the quantum geometry can only be determined by the underlying - and yet unknown - QG theory. Therefore we have to make do with what we have and make assumptions about the kind of states that could be generated by a reasonable QG theory. The uncertainty relations of $\kappa$-Minkowski, $\Delta \hat z^0 \Delta \hat z^i \geq  \frac \kappa 2 \langle \hat z^i \rangle$, allow for states with arbitrarily small spatial or temporal uncertainty, but unless the spatial coordinate $\hat z^i$ is centred on the origin ($\langle \hat z^i \rangle = 0$), this comes at the cost of a correspondingly large uncertainty in the other coordinate. We assume that a `reasonable' QG theory won't predict states with a macroscopically large uncertainty in any of the coordinates. Instead, it is expected to predict a quantum geometry that is well approximated by the commutative Minkowski space. This means that we should be looking for states in which the squared sum of uncertainties of all coordinates, $(\Delta \hat z^0)^2 + \sum_i(\Delta \hat z^i)^2$ is near its minimum $\frac \kappa 2 \langle \hat z^i \rangle$. We call these states `semiclassical'.
As we remarked, without knowing the details of the underlying QG theory we have no way of predicting the details of the wavefunction, but we expect the essential features of the expectation value of the Pauli--Jordan function to be independent of these details, as long as the state satisfies the semiclassicality condition. To test this hypothesis we calculate $\langle \hat{\Delta}_\text{PJ}(\hat x,\hat y)\rangle $ on four different wavefunctions satisfying the semiclassicality condition, and verify that the results are indistinguishable from one another.

To define our Hilbert space $\mathcal H$ of `geometric' states we introduce a representation of the algebra~\eqref{k-Minkowski_algebra}. $\hat z^i$ will act as multiplicative operators on a space of complex functions $\psi(\vec q)$ on $\mathbbm{R}^n$ (where $n$ is the number of spatial dimensions), while $\hat z^0$ is represented as a differential operator  
\begin{equation}\label{kappa-Representation}
\hat z^i  \psi(\vec q) = \frac{\vec q}{\kappa} \psi(\vec q) \,, ~ \hat z^0   \psi(\vec q) = - i \vec q \cdot \frac{\partial \psi(\vec q)}{\partial \vec q} - i \frac{n}{2}  \psi(\vec q) \,.
\end{equation}
We then define an inner product on $\mathcal H$ as $
\langle \varphi  | \psi \rangle = \int_{\mathbbm{R}^n} d^n q ~ \bar \varphi(\vec q) \psi(\vec q) \,.
$
These definitions give a faithful representation of the algebra~\eqref{k-Minkowski_algebra}.

From now on, for simplicity, we will work in 1+1 dimensions ($n=1$). Consider a Gaussian:
\begin{equation}\label{GaussianWavepacket}
\gamma(q ; \bar z^0,\bar z^1) = \left( \frac 2 {\sigma \sqrt{\pi}} \right)^{\frac 1 2} e^{- \frac{(q-\kappa \bar  z^1 )^2}{\sigma^2}} e^{i \frac{\bar  z^0}{\bar  z^1} q } \,.
\end{equation}
The above wavefunction has $\langle \gamma | \hat z^\mu | \gamma \rangle =\bar  z^\mu$, and the squared sum of uncertainties $(\Delta \hat z^0)^2+(\Delta \hat z^1)^2$ is minimized\footnote{The minimum is $\frac 1 2 \kappa^{-2}+ \kappa^{-1}\sqrt{(\bar z^0)^2 + (\bar z^1)^2}$.} if we fix the parameter $\sigma$ to the value $\sigma = 2 \kappa^{\frac 1 2} \bar  z^1 [(\bar z^0)^2 + (\bar z^1)^2]^{-\frac 1 4}$. We now want to work with two points, and therefore we generalize the representation~\eqref{kappa-Representation} to the  tensor-product algebra $\mathcal A \otimes \mathcal A$: $\hat x^1 = \frac{q_x}{\kappa}$, $\hat x^0  =- i  q_x \partial_{q_x} - \frac{i}{2}$, $\hat y^1 = \frac{q_y}{\kappa}$, $\hat y^0  = - i q_y \partial_{q_y} - \frac{i}{2} $. 

We can then define a semiclassical state of the two coordinates $\hat x$ and $\hat y$ as $\psi(q_x,q_y) = \gamma(q_x,\bar x^0,\bar x^1)\gamma(q_y,\bar y^0,\bar y^1)$, which has expectation values $\langle \psi | \hat x^\mu | \psi \rangle =\bar  x^\mu$,  $\langle \psi | \hat y^\mu | \psi \rangle =\bar  y^\mu$, and minimal squared sums of uncertainties for both points. To calculate the expectation value of the Pauli--Jordan function $\langle \hat{\Delta}_\text{PJ} \rangle$ on this state it is then sufficient to observe that the exponentials in~\eqref{Noncommutative_Pauli-Jordan_3D} act on our wavefunction as:
\begin{equation}
\begin{aligned}
e^{i p_1  \hat x^1 - i e^{\frac{\omega^\pm }\kappa} p_1  \hat y^1  }  e^{i \omega^\pm (\hat x^0 - \hat y^0)}  \psi(q_x,q_y)
=
\\
e^{i \frac{p_1 q_x^1}{\kappa} - i e^{\frac{\omega^\pm }\kappa} \frac{p_1 q_y^1}{\kappa}  }   \psi(e^{-\omega^\pm}q_x,e^{+\omega^\pm}q_y) \,,
\end{aligned}
\end{equation}
and sandwich the expression above with $\bar \psi(q_x,q_y ; \bar  x^\mu,\bar  y^\mu)$, integrating in $\diff q_x \diff q_y$ over all of $\mathbbm{R}^2$:
\begin{equation}\label{Noncommutative_Pauli-Jordan_1D}
\begin{aligned}
\langle \hat{\Delta}_\text{PJ} \rangle 
&= \int_{\mathbbm{R}} \!\!\! \diff p_1 
 \frac{
\text{sign}(\kappa-|\vec p|)}{ 2 \sqrt{ \left(m^2+| \vec p |^2\right)}}
\int_{\mathbbm{R}^2} \!\!\! \diff q_x \diff q_y ~ F(\omega^+)
\\
&-
\int_{\mathbbm{R}}\!\!\!  \diff p_1 \frac{1}{ 2 \sqrt{ \left(m^2+| \vec p |^2\right)}}  \int_{\mathbbm{R}^2}   \!\!\! \diff q_x \diff q_y ~ F(\omega^-)
\,,
\end{aligned}
\end{equation}
where
\begin{equation}
F(\omega) = 
e^{i \frac{p_1 q_x^1 - i e^{\frac{\omega  }\kappa} p_1 q_y^1}{\kappa}  } 
\bar \psi(q_x,q_y)  \psi(e^{-\omega }q_x,e^{+\omega }q_y) \,.
\end{equation}
The integral in $\diff q_x \diff q_y$ of $F(\omega)$ can be easily performed analytically, and the result inserted into expression~\eqref{Noncommutative_Pauli-Jordan_3D} gives an expression that can be integrated in $\diff p_1$ numerically. The result is an ordinary, commutative function $\langle \hat{\Delta}_\text{PJ} \rangle =\Delta^\kappa_\text{PJ}(\bar x,\bar y) $ of only the (numerical) expectation values of the coordinates, $\bar x^\mu$ and $\bar y^\mu$.

\begin{figure}[t!]\center
\includegraphics[width=0.48\textwidth]{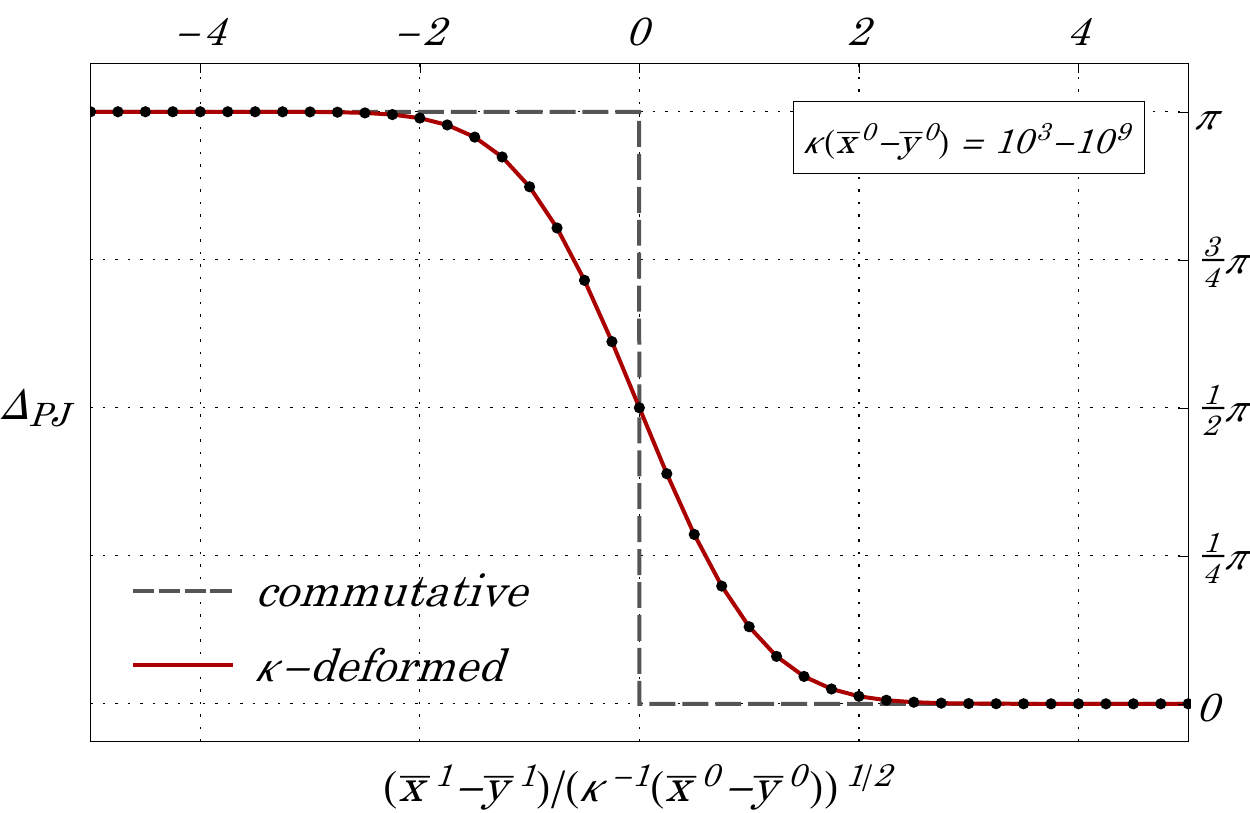}
\caption{Cross-section of $\Delta^\kappa_\text{PJ}(\bar x,\bar y)$ near the light-cone (in red, black dots represent numerical sampling points), vs. its commutative counterpart. We fixed the time interval $\delta \bar x^0$ and plotted  $\Delta^\kappa_\text{PJ}(\bar x,\bar y)$ for an interval of values of $\delta \bar x^1 $. Rescaling the horizontal axis by $\sqrt{\kappa^{-1} \delta x^0}$, the form of the cross-section does not depend on $\delta \bar x^0$.
}
\label{Pauli-Jordan_fig}
\end{figure}

\begin{figure}[t!]\center
\includegraphics[width=0.48\textwidth]{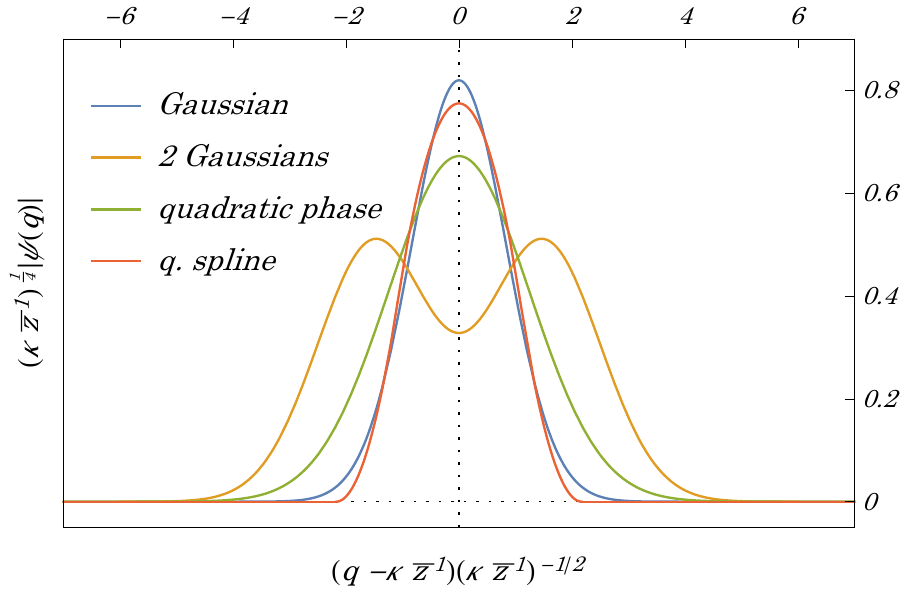}
\caption{Absolute value of the four test-wavefunctions in the vicinity of the light-cone $\kappa \bar z^0 = \kappa \bar z^1$. The horizontal axis has been rescaled  by $\sqrt{\kappa \bar z^1}$, and the vertical axis by $(\kappa \bar z^1)^1/4$. This makes the waveforms independent of the choice of $\bar z^1$.
It is apparent here that the results shown in Fiq.~\ref{Pauli-Jordan_fig} do not depend on the details of the wavefunction, as long as it is `semiclassical', \emph{i.e.} localized in a region whose size makes both the space and the time uncertainties of the same order of  magnitude.}
\label{Test_Functions_fig}
\end{figure}

$\Delta^\kappa_\text{PJ}$ is indistinguishable from the commutative Pauli--Jordan function everywhere except close to the classical light-cone $\delta \bar x^0 = \delta \bar x^1$ (where $\delta \bar x^\mu = \bar x^\mu - \bar y^\mu$). In that region, instead of abruptly jumping from the value $\pi \, \text{sign} (\delta \bar x^0)$ for $\delta  \bar x^0 > \delta \bar x^1$ to the value $0$ when $\delta \bar x^0 < \delta \bar x^1$, our function smoothly interpolates between these two values over a region of size $\mathcal{O}[\sqrt{\kappa^{-1}\delta \bar x^0}]$ (the geometric mean between the Planck length $\kappa^{-1}$ and the distance from the origin). The form of the interpolating function is exactly the same at any distance $\delta \bar x^0$ from the origin, if we express the spatial interval $\delta  \bar x^1$ in units of $\sqrt{\kappa^{-1} \delta \bar x^0 }$ (see Fig.~\ref{Pauli-Jordan_fig}). In other words, the function shows an \emph{anisotropic scale invariance} under the transformation $\delta \bar x^1 \to \alpha \delta \bar x^1$, $\delta \bar x^0 \to \alpha^2 \delta \bar x^0$ [as do the $\kappa$-Minkowski commutation relations~\eqref{k-Minkowski_algebra}].

 We checked this over an interval of values of $\delta \bar x^0$ between $10^3$ and $10^9$ Planck times, as large as our numerical accuracy allowed us. Realistic values for the parameters ($\delta \bar x^0 \sim 10^9 ly$ and $\kappa^{-1} \sim 10^{-44} s$) would yield a value of $10^{59}$ Planck times, far beyond what can be explicitly checked at the calculator; however the anisotropic scaling property we showed in the interval $10^3-10^9$ Planck times provides sufficient evidence that the form of the function is the same also at realistic scales.

In~\cite{Weblink}, available online, we produced an animation showing $\Delta^\kappa_\text{PJ}$  in the plane $\delta \bar x^1 - \delta \bar x^0$, zooming away from the origin. Figure~\ref{LightCone2D_1} shows three stills from the animation. This shows how $\Delta^\kappa_\text{PJ}$  becomes indistinguishable from the commutative Pauli--Jordan function at large scales. Moreover, compare our animation with the Figures of~\cite{Brasiliani}, where the `blurry' region has the same thickness everywhere; the boost in the size of our nonlocal effect is apparent.

Finally, we studied the dependence of the above results from the details of the wavefunction $\psi$, choosing a few sufficiently-varied examples among the `semiclassical' wavefunctions. We chose three examples that have the good property of being analytically integrable (except for the last step of the integral in  $\diff p_1$). One is the sum of two Gaussian wavepackets peaked at $q = \kappa \bar z^1 \pm \frac a 2  \sqrt{\kappa \bar z^1}$, with variance $\sigma = \sqrt{\kappa \bar z^1}$, both multiplied by the same phase $\exp( i \frac{\bar x^0}{\bar x^1} q)$ (and of course normalized together). If $a$ is an $\mathcal{O}(1)$ number the squared sum of the variances $(\Delta \hat z^0)^2 + (\Delta \hat z^1)^2$ is still near its minimum, and the two wavepackets can be clearly distinguished, producing a `double-peaked' waveform (see Fig.~\ref{Test_Functions_fig}). Yet, the dependence of the expectation value $\langle \hat{\Delta}_\text{PJ}  \rangle$  on $\bar x^\mu$ and $\bar y^\mu$ is indistinguishable from the one shown in Fig.~\eqref{Pauli-Jordan_fig}.
The second example we checked was to modify the Gaussian~\eqref{GaussianWavepacket} by multiplying it by an imaginary-variance Gaussian term $e^{i (c \, q^2 + d \, q)}$, and adjusting the parameters to satisfy the semiclassicality requirement. The third example was to use a quadratic spline forming a twice-differentiable (necessary to calculate the variances) symmetric bell-shaped curve, with a compact support. The normalization condition leaves only one free parameter for such a function, which is then fixed by the semiclassicality condition. None of those examples gives rise to an appreciable modification to the Pauli--Jordan expectation value.

\section{Conclusions}

We studied the Pauli--Jordan function for a scalar QFT on $\kappa$-Minkowski spacetime. This defines the commutator of fields at different points, and consequently their causality relations. The standard formulation of microcausality in quantum field theory is that all observables should commute at spacelike-separated points.

We listed two objectives in the introduction, which our analysis fulfilled completely:
\begin{enumerate}
\item  The claim of Ref.~\cite{Brasiliani} about the size of the nonlocal effects in $\kappa$-Minkowski is rejected by our analysis. Rather than an in-principle untestable Planckian fuzzyness, 
our $\kappa$-deformed Pauli--Jordan function is nonzero outside of the light-cone over a region of size $\sim \sqrt{L_p~L}$, \emph{i.e.}, the geometric average of the Planck length and the distance from the origin. For cosmological distances ($\sim 10^9 ly$) this amounts to an effect \emph{30 orders of magnitude larger}.

\item Our analysis, we believe, casts serious doubts on the possibility of superluminal propagation in $\kappa$-Minkowski. We emphasize again that our work relies on a few plausible, but yet unproven, assumptions. In particular we assume that only the `semiclassical' states on the $\kappa$-Minkowski algebra which we defined above are relevant to our discussion. Another key assumption was that the Pauli--Jordan function still entails the causal structure of fields on a noncommutative spacetime.

Our $\kappa$-deformed Pauli--Jordan function is  nonzero outside of the light-cone, but only over a region of size $\sim \sqrt{L_p~L}$. This kind of nonlocality is compatible with the uncertainty relations predicted by~\eqref{k-Minkowski_algebra} and can therefore be attributed to the intrinsic fuzzyness of spacetime points.
This result effectively excludes the possibility of constraining our models with GRBs. In fact, the present sensitivities are somewhere $14$ orders of magnitude worse than what would be needed to measure the kind of superluminal propagation we predict. 

\end{enumerate}

We stress that, in principle, the lightcone we found allows signals from distant sources to be detected with a spacetime spread  much larger than the Planck scale. Our calculation allows us to make an educated guess for what regards the size of the expected effect: a pointlike source placed at a distance $L$ from the detector will be detected with a spread in space and time of order $\sim\sqrt{L_p~L}$. For a distant GRB $L \sim 10^9 ly$, and therefore $\sqrt{L_p~L} \sim 10^{-14} s$. This would be measurable with present-day technology if we had at our disposal an ultrashort-duration source, \emph{e.g.} a femtosecond laser, placed 1 billion light years away from Earth. Unfortunately we presently have no knowledge of such ultra-localized sources in our universe.

It is interesting also to compare our result with past work on the implications of Quantum Gravity for the light cone. In particular, refs.~\cite{Ford0, Ford1} describe some simple models in which a (linearized) graviton background induces quantum fluctuations of the lightcone. The results depend on the assumed state of the graviton background, and there is a particular choice which reproduces the distance-dependence of our result. Namely, when the gravitons are thermal and their typical wavelengths are much smaller than the distance at which the signal is detected. 
In this case, the fuzzyness in time is proportional to the square root of the distance from the source of the signal, times a quasi-constant logarithmic correction: $\Delta t \sim\frac{L_p}{c} \sqrt{\frac 3 {180\pi}  L \, T \left[ \log \left( L \, T \right) + c_1 \right]}$, where $c_1$ is a constant of order unity. The proportionality factor is
 $\sqrt{L_p \, T\,  \log (LT) / 180}$, which, for temperatures $T$ comparable to that of the CMB, gives a suppressing factor of the order of $10^{-16}$ with respect to the effect we predicted. To get an effect of similar magnitude the thermal bath should have Planckian temperature.

\section*{Acknowledgements}

F.M. was funded by the European Union and the Istituto Italiano di Alta Matematica under a  Marie Curie COFUND action.


\providecommand{\href}[2]{#2}\begingroup\raggedright\endgroup

\end{document}